\documentclass[preprint2]{aastex}
\pdfoutput=1
\usepackage{natbib}
\usepackage{graphicx}

\shorttitle{Perpendicular transport of Solar Energetic Particles}
\shortauthors{M.S. Marsh et al.}

\begin{document}

\title{Drift-induced perpendicular transport of Solar Energetic Particles}

\author{M. S. Marsh, S. Dalla, J. Kelly, and T. Laitinen}
\affil{Jeremiah Horrocks Institute, University of Central Lancashire, Preston, PR1
2HE, UK}
\email{mike.s.marsh@gmail.com}

\begin{abstract}
Drifts are known to play a role in galactic cosmic ray transport within the heliosphere and are a standard component of cosmic ray propagation models. However, the current paradigm of Solar Energetic Particle (SEP) propagation holds the effects of drifts to be negligible, and they are not accounted for in most current SEP modelling efforts.
We present full-orbit test particle simulations of SEP propagation in a Parker spiral interplanetary magnetic field which demonstrate that high energy particle drifts cause significant asymmetric propagation perpendicular to the interplanetary magnetic field. Thus in many cases the assumption of field aligned propagation of SEPs may not be valid. We show that SEP drifts have dependencies on energy, heliographic latitude, and charge to mass ratio, that are capable of transporting energetic particles perpendicular to the field over significant distances within interplanetary space, e.g. protons of initial energy 100 MeV propagate distances across the field on the order of 1 AU, over timescales typical of a gradual SEP event. Our results demonstrate the need for current models of SEP events to include the effects of particle drift. We show that the drift is considerably stronger for heavy ion SEPs due to their larger mass to charge ratio. This paradigm shift has important consequences for the modelling of SEP events and is crucial to the understanding and interpretation of in-situ observations.
\end{abstract}


\section{Introduction}\label{sect_intro}
Magnetically triggered eruptive events in the solar atmosphere, such as coronal mass ejections (CMEs) and flares, cause the acceleration and ejection of a large swathe of energetic ions and electrons into the interplanetary magnetic field environment. These solar energetic particle (SEP) events lead to a sudden, transient, increase in the flux of ionising radiation along their locus of propagation within the heliosphere. Such events affect the local space environment and, in particular, can impact human activities such as satellite technology, the biological risks of human spaceflight, and even the terrestrial radiation environment; these events form part of a field of study collectively termed space weather. 

An increased reliance on space technology systems and planned further human exploration of space has led to a pressing demand for the transition from research models of space weather events to actionable operational models that may be used to mitigate against the impact of space weather. It is therefore important that any operational model includes an accurate description of the important physical processes that determine the parameters of an event. 

There has been much effort from the scientific community in developing research models of SEP propagation in the interplanetary magnetic field (IMF). Most of the work in this field has adopted a description based on the focussed transport equation, originally developed by \cite{Roe1969} and further refined in a number of works \citep[e.g.][]{Ruf1995}. This approach underpins a number of studies including models that incorporate acceleration at a propagating interplanetary shock \citep[e.g.][]{Lar1998}.  
Historically, the focussed transport equation is based on the axiom that propagation perpendicular to the interplanetary magnetic field is negligible. This implies that there has to be a direct magnetic connection between the particle's source region and the location in interplanetary space where the SEP event is measured.

SEPs have been detected by the Solar Terrestrial Relations Observatory (STEREO) spacecraft at locations widely separated in longitude \citep[][]{Dre2012}, confirming earlier observations made by Helios. The STEREO data have also shown that particles in so-called impulsive events, thought to be associated with a localised source region at the Sun, can propagate across the interplanetary magnetic field \citep{wie13}. 
Observations have also shown that the flux profiles of large SEP events measured at different locations within the inner heliosphere become homogeneous \citep{mck72}, described as a particle reservoir effect,  implying a smoothing of the azimuthal gradient of the event flux. 
As discussed in \cite{wie13}, a number of possible scenarios that may account for a wide longitudinal extent of SEP events have been postulated:  large values of the ratio of diffusion coefficients perpendicular and parallel to the mean magnetic field, divergence and braiding of the field between the photospheric footpoints and solar wind source surface, and global scale shock acceleration and reconfiguration of the heliospheric field due to CMEs. Thus, although it is now generally accepted that some transport across the mean field is taking place, the mechanisms responsible remain unclear. 

Recently, a focussed transport equation that includes propagation across the field via  a perpendicular diffusion coefficient has been introduced  and solved numerically by means of a stochastic differential equation (SDE) method \citep{Zha2009}. This approach has been applied to the analysis of SEP transport from a localised region at the Sun \citep{Zha2009, Dro2010} and from an extended CME shock \citep{Wan2012}. Within these models it is assumed that propagation across the field is symmetric with respect to the magnetic field.
However, most current numerical codes aiming to predict SEP fluxes for space weather applications neglect transport perpendicular to the magnetic field \citep[e.g.][]{Ara2006, Luh2010}. The energetic particle module within the EMMREM model \citep{Sch2010} accounts for cross field transport by prescribing diffusion coefficients and including a drift term that is averaged over pitch angle.

The full-orbit test particle method offers an alternative approach to the solution of transport equations for the modelling of SEP propagation. The advantage of this method is that it does not require simplifying assumptions to reduce the number of variables in the problem, since the physics of particle propagation is determined by the solution of the equations of motion alone. It is therefore ideally suited to studying transport across the magnetic field. \cite{Pei2006} and \cite{kel12} used this method to analyse transport in a Parker spiral magnetic field with large scale fluctuations, and \cite{tau11} investigated the effect of small scale turbulence. 

Particle drifts due to gradients and curvature in the large scale Parker spiral IMF are known to be important in the propagation of galactic cosmic rays (GCRs) \citep{Jok1977} and are included in standard GCR models based on the Parker transport equation \citep[e.g.][]{Fer2004}. 
In the current paradigm for SEPs, drifts have been considered unimportant and, up to the present time, neglected in most propagation models. 

Two early studies of SEP drifts in the Parker spiral interplanetary magnetic field exist \citep{bur68, Win1968}, both of which conclude that the particle motion traces out the magnetic field lines on which they were originally injected and drifts are negligible. Recently, \cite{anal_pap} carried out a reanalysis of the analytical expressions for drifts in the Parker spiral and pointed out that the magnitude of drift velocities can be significant for particles at the high energy end of the SEP range.

In this paper, SEP propagation is modelled numerically using a full-orbit test particle code. The modelling approach and simulations are discussed in Section~\ref{sec.model}. It is found that drifts can be an important cause of perpendicular propagation of SEPs, particularly for the energy ranges that have space weather impact. These results are described in Section~\ref{sec.results} and discussed in Section~\ref{sec.conclusions}.

\begin{figure}[t]
%\epsscale{1.0}
%\epsscale{0.75}
\centering
%\rotate
\plotone{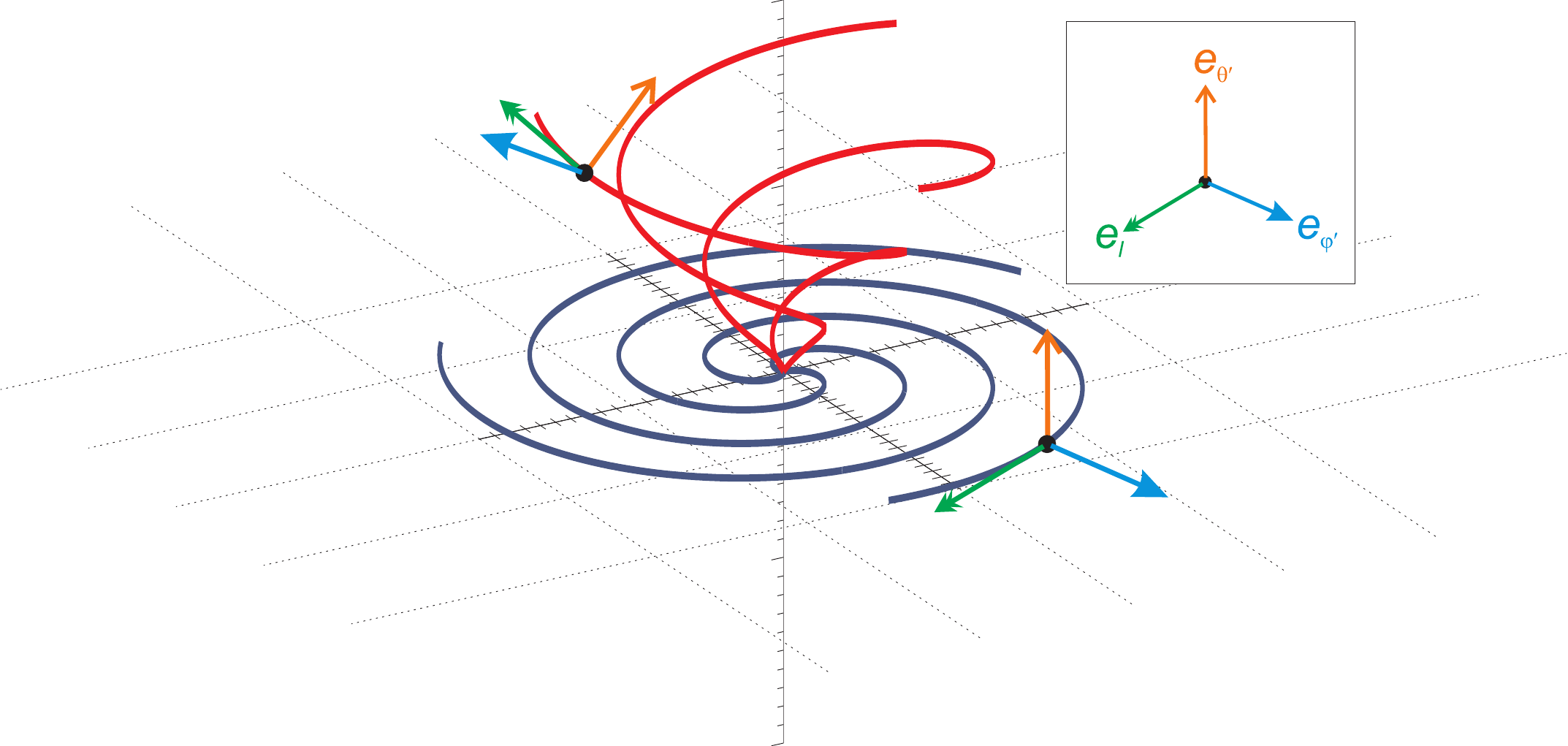}
\caption{
Schematic diagram illustrating the local spiral coordinate system with Parker field lines plotted at 0 and 45$^{\circ}$ latitude (blue and red curves respectively). Two example target points (black circles) show the relative orientations of the local spiral coordinate system axes at two different locations within the IMF. The legend indicates the arrow type and color representing each axis where $\mathbf{e}_{l}$ (green), $\mathbf{e}_{\theta}$ (orange), and $\mathbf{e}_{\phi'}$ (blue).}
\label{local_coord}
\end{figure}

\section{Modelling} \label{sec.model}
The propagation of energetic particles is modelled using a relativistic full-orbit test particle numerical code, originally developed to study particle acceleration due to magnetic reconnection \citep{dall05}. 
The equations of motion of the particles are solved numerically using an adaptive-step Bulirsch--Stoer method \citep{pre96}. The numerical code has been adapted to study charged particle propagation in the heliosphere, and recently applied to the investigation of the effect of magnetic turbulence on particle propagation \citep{kel12, lai12}. 

\subsection{Interplanetary magnetic and electric fields}
Particle trajectories are calculated in a Parker spiral interplanetary magnetic field, in a heliocentric non-rotating fixed reference frame, assuming a solar wind flow that is radial, uniform, and time independent. 
To illustrate the magnitude and effect of drifts on SEPs, a simple unipolar Parker magnetic field is used in our simulations without the presence of a current sheet.
The magnetic field is given by:
\begin{equation}\label{eqn_1}
\mathbf{B} =  \frac{B_{0} r_{0}^{2}}{r^{2}} \, \mathbf{e}_{r} - \frac{B_0  \, r_0^2 \, \Omega}{v_{sw}} \, \frac{\sin{\theta}}{r} \, \mathbf{e_{\phi}},
\end{equation}
where $(r, \theta, \phi)$ are spherical coordinates giving radial distance, colatitude and longitude respectively, $B_{0}$ is the radial component of the magnetic field strength at the reference surface of radius $r_{0}$,  $\Omega$ is the solar angular rotation rate (assumed constant), $v_{sw}$ is the solar wind speed and the $\mathbf{e}$ parameters represent the unit vectors.
 
In the fixed reference frame, in which the solar wind is moving radially outward at speed $v_{sw}$, an electric field $\mathbf{E}$ = $-$ $ \mathbf{v}_{sw} /c \times \mathbf{B}$ is present, which, using Equation~(\ref{eqn_1}), takes the form:
\begin{equation}\label{eqn_2}
\mathbf{E} = - \frac{\Omega  \, B_0 r_0^2}{c}   \,  \frac{\sin{\theta}}{r} \, \mathbf{e_{\theta}},
\end{equation}
where $c$ is the speed of light.

Associated with this electric field is a corotation drift that ensures particles follow the rotation of interplanetary magnetic flux with time. 

In our simulations we use $B_{0}$=178 $\mu$T, $r_{0}$=$6.96 \times 10^{5}$~km (one solar radius), $\Omega$
=$2.86 \times 10^{-6}$ rad s$^{-1}$ and $v_{sw}$=$500$~km~s$^{-1}$.
The values of $B_{0}$ and $r_{0}$ ensure that the magnetic field magnitude at 1 AU is 5~nT.
Equation~(\ref{eqn_1}) describes a positive unipolar field, i.e.~the presence of two polarities and a current sheet is not included. 

Equations~(\ref{eqn_1}) and (\ref{eqn_2}) are substituted into the particle's equation of motion:
\begin{equation}  \label{eqn_3}
\frac{d {\bf p}}{d t}  = q \, \left( {\bf E} + \frac{1}{c}  \frac{{\bf p}}{m_0 \gamma} \times {\bf B} \right),
\end{equation}
where ${\bf p}$ is the particle's momentum, $t$ is time, $q$ its charge, $m_0$ its rest mass and $\gamma$ its Lorentz factor. Equation~(\ref{eqn_3}) is numerically integrated for each particle in a population.
  
The precision of the code is tested by ensuring conservation of the total particle kinetic and potential energy under scatter free conditions. 
\subsection{Scattering}
The effect of particle interaction with small-scale interplanetary turbulence is simulated by introducing random scattering of the particle's velocity \cite[c.f.][]{Pei2006,kel12}. Isotropic scattering is implemented in the solar wind reference frame, describing scattering due to magnetic turbulence embedded within the solar wind. 
The level of scattering is determined by a prescribed mean free path $\lambda$. The scattering events are Poisson distributed in time with an average scattering time $t_{scat}= \lambda / v_0$, where $v_0$ is the initial particle velocity.
The scattering inter-event waiting times are therefore exponentially distributed allowing the particle's equations of motion to be integrated up to the next scattering event where the direction of the particle's velocity vector is randomly reassigned and the integration then proceeds.

\subsection{Simulation initial conditions}

We inject populations of particles with initial positions randomly distributed in a region of 8$^{\circ}$$\times$8$^{\circ}$ in longitude and latitude at 1 solar radius. In each simulation the initial injection energy $E_0$ is the same within the particle population and velocity vectors are randomly distributed in the semi-hemisphere of pitch angles between 0 and 90$^{\circ}$ (i.e. with velocities outward from the Sun). 
The particles within a population are injected simultaneously at initial time $t=0$, the simulation boundary is open, and all particles are conserved throughout each simulation.

\subsection{Local spiral coordinate system and calculation of displacements}\label{sec.transform}

To quantify the degree of transport across the field, the perpendicular displacement of each particle from the Parker field line on which it was originally injected is calculated (in the following, this field line will be referred to as the initial field line). 
If $P$ is the particle position at a given time, the vector from $P$ that intersects the initial field line perpendicularly defines the target point position on the initial field line $P_{t}$. The magnitude of the displacement is then $\Delta s = |P - P_{t}|$.
The target point is used to define the origin of a local Parker spiral coordinate system used for the calculation of the displacements $\Delta s$ \citep[see][]{kel12,tau11}.
The local coordinate system $(\mathbf{e}_{l},\mathbf{e}_{\phi'},\mathbf{e}_{\theta'})$ has an axis $\mathbf{e}_{l}$ that is tangential to the magnetic field line vector at the target point and
directed outwards, another axis in the direction of $\mathbf{e}_{\theta'}$=$-$$\mathbf{e}_{\theta}$ (with $\mathbf{e}_{\theta}$ the standard spherical coordinate system unit vector) that is perpendicular to the surface of constant latitude containing the initial field line, and an axis $\mathbf{e}_{\phi'}$  completing the right-handed orthogonal system as described in \cite{kel12}. Figure~\ref{local_coord} illustrates the relative orientation of the axes originating at two example target points located at different points in the IMF. This local coordinate system is used to calculate the perpendicular displacement of each particle from its target point in the lateral and latitudinal directions, $\Delta s_{\phi'}$ and $\Delta s_{\theta'}$ respectively. 

\begin{figure}[!t]
%\epsscale{1.0}
\epsscale{0.77}
\centering
%\rotate
\plotone{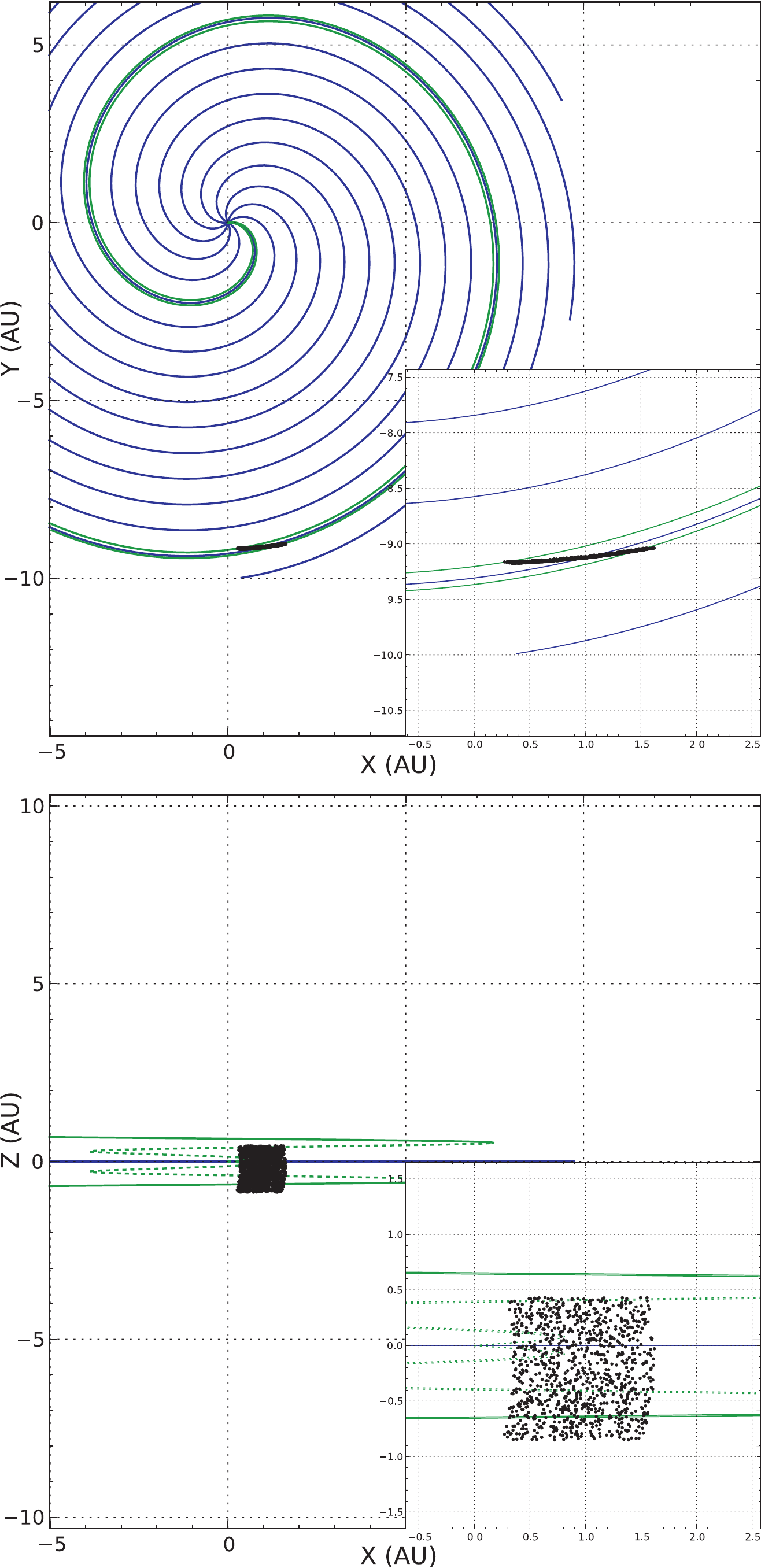}
\caption{
The ($x$-$y$) (top) and ($x$-$z$) (bottom) locations of 1000 protons, with injection energy $E_{0}$=100 MeV, after scatter-free propagation, at time $t$=12~hrs. Particles are indicated by black dots. Blue curves show the equatorial IMF, green curves show the bounding magnetic field lines originating from the corners of the injection region centered at latitude $\delta_{0}$=0 (in the lower panel the inner spiral region of the bounding lines is dashed and the outer region, where the particles are located, is indicated as the solid green lines). Each plot inset shows an enlarged region around the particle locations. Note the displacement of the particles perpendicular to the equatorial plane in the ($x$-$z$) projection. }
\label{scat_free}
\end{figure}

\section{Results}\label{sec.results}

In the following, we conduct a parametric study of drifts in the interplanetary magnetic field, varying the scattering mean free path $\lambda$, the latitude $\delta_{0}$ of the centre of the injection region, and the initial particle injection energy $E_0$.

\subsection{Scatter-free propagation}
%Mean energy corresponding to plot (spec12) E=94 MeV, energy at spec100 E=84.

Figure~\ref{scat_free} shows the the spatial distributions of $E_0$=100~MeV protons at a time $t$=12~hrs for the case when no scattering is taking place. The number of particles in the population is $N$=1000 and the injection region is centered at heliographic latitude $\delta_{0}$=0$^{\circ}$  (in the following the degree symbol will be omitted for latitudes).

The top panel shows an ($x$-$y$) projection and the bottom panel an ($x$-$z$) projection. Here the $z$ axis is the rotational axis of the Sun, while the $x$ and $y$ coordinates lie in the heliospheric equatorial plane with the $x$ axis corresponding to a longitude of $\phi=0^{\circ}$.
In all panels, the green curves are Parker spiral magnetic field lines with starting points at the four corners of the particle injection region and the blue curves show the IMF in the heliospheric equatorial plane. The plotted field lines are rotated with respect to their location at $t=0$ to account for solar rotation. Particles are indicated by black dots and the insets focus on their location in more detail.

\begin{figure*}[!ht]
\epsscale{2.2}
%\epsscale{0.65}
\centering
%\rotate
\plotone{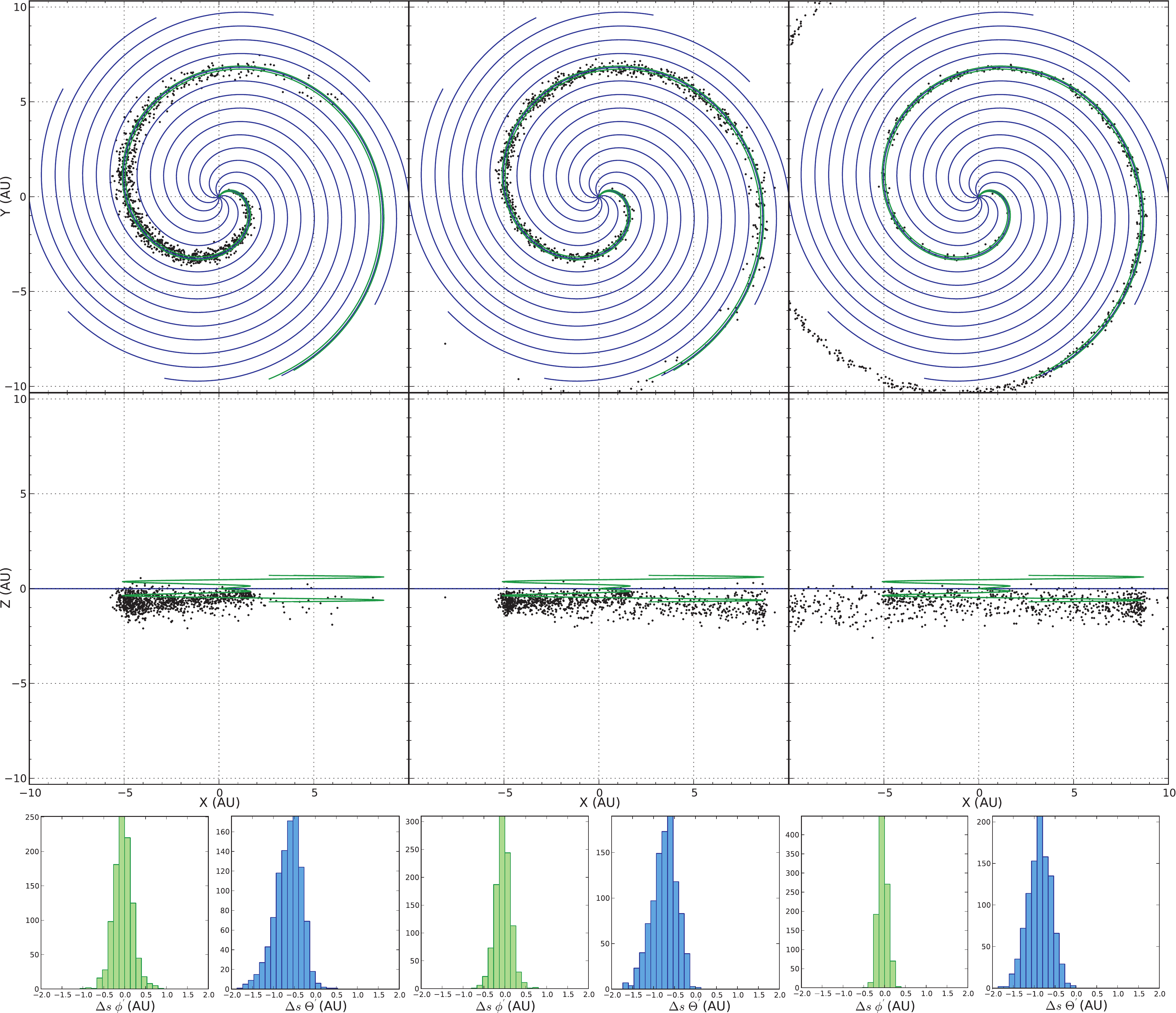}
\caption{Locations of 1000 protons with $E_0$=100 MeV at $t$=4~days for $\lambda$=0.3 (left column), 1 (middle column), and 10 AU (right column), for injection at $\delta_{0}$=0. Top and middle panels give the ($x$-$y$) and ($x$-$z$) projections, while the bottom panels show histograms of displacements $\Delta s_{\phi'}$ (green) and $\Delta s_{\theta'}$ (blue) from the initial field line. In the top and middle panels, blue curves show the equatorial IMF with the field lines bounding the particle injection region in green. The online edition of the paper includes a movie showing the case of $\lambda=$1~AU. }
\label{100MeV_3scat}
\end{figure*}

In the scatter-free case, particles are very quickly focussed by the decrease in the magnitude of $\mathbf{B}$ in the Parker spiral. 
Figure~\ref{scat_free} shows that the particles propagate closely together with similar velocity along the magnetic field, describing a sheared surface. In the ($x$-$y$) projection, the surface described by the original injection region is sheared, due to the geometrical effect of propagation along the spiral. Since each particle travels the same total distance, particles within the distribution located on field lines forming leading parts of the spiral tend to lag behind the rest of the distribution.
In the ($x$-$z$) projection, the surface is also sheared due the variation in curvature of the Parker field lines in the azimuthal direction as a function of latitude. The field lines originating at higher latitudes have a smaller radius of curvature in the azimuthal direction, giving the effect that particles on these field lines lead the rest of the distribution as they propagate outwards along the spiral.
In the latitudinal direction, a systematic displacement perpendicular to the equatorial plane is observed in the ($x$-$z$) projection: the inset in the lower panel of Figure~\ref{scat_free} clearly shows that the particles are no longer within the green lines delimiting the injection region.

\subsection{Propagation with scattering}
%mean energy for plots in Fig. L=0.3 E=53 MeV, L=1 E=64, L=10 E=78

We now examine the propagation of 100 MeV protons injected with the same initial parameters as in Figure~\ref{scat_free}, into an IMF in which scattering is present,
and at time $t$=4~days to allow the distance travelled to be comparable to the scatter free case. 
Three values of the scattering mean free path $\lambda$ are considered: $\lambda$=0.3, 1 and 10~AU. These represent very different propagation conditions:  $\lambda$=0.3~AU is considered a relatively high scattering regime, and is of the order that is obtained when measured proton profiles for gradual SEP events are fitted by means of a focussed transport model \citep[e.g.][]{kal97}. Recently, it has been argued that the value of $\lambda$ is in fact considerably larger and likely close to 1 AU (Reames 1999). The value $\lambda$=10~AU gives a very low scattering condition.

Figure~\ref{100MeV_3scat} shows the location of the protons at time $t$=4~days for $\lambda$=0.3~AU (left column), 1~AU (middle column) and 10~AU (right column). The first two rows show the ($x$-$y$) and ($x$-$z$) projections and the bottom row shows histograms of the perpendicular displacement from the field line on which each particle was injected, calculated as described in Section~\ref{sec.transform}. 

As expected, the extent to which the particle distribution is able to propagate along the field is dependent on the mean free path. 
The notable feature of Figure~\ref{100MeV_3scat} is that in all 3 scattering conditions, particles do not remain tied to their original field line but spread perpendicular to the field. The population is not confined within the bounding field lines (green curves) originating from the corners of the injection region. The displacement of the particles can be categorized into the two components defined by $\mathbf{e}_{\phi'}$ and $\mathbf{e}_{\theta'}$ as a lateral and latitudinal displacement.
The latitudinal displacement is visible in the ($x$-$z$) projections as a displacement in the direction perpendicular to the plane described by the field line on which the particle is located; in the special case around the equatorial plane, this is approximately along the negative $z$ axis, in a similar manner to that shown in Figure~\ref{scat_free}. The lateral displacement is visible in the ($x$-$y$) projections as a dispersal outward beyond the field lines bounding the injection region.

The bottom panels of 
Figure~\ref{100MeV_3scat} show histograms of the displacements in the $\phi'$ and $\theta'$ directions which we define as the lateral displacement $\Delta s_{\phi'}$ and the latitudinal displacement $\Delta s_{\theta'}$. The width of the $\Delta s_{\theta'}$ histogram appears similar in the three scattering conditions of  Figure~\ref{100MeV_3scat} with a slightly decreasing width with $\lambda$. 
It can also be seen that the peak of the $\Delta s_{\theta'}$ distribution shifts to increasing magnitudes of displacement with increasing $\lambda$. This can be understood in the following section where Equations~\ref{eq.vdphipr} and \ref{eq.vdthetapr} show that less scattering (i.e. larger $ v_{\|}$) results in a maximum drift velocity magnitude.
The lateral $\Delta s_{\phi'}$ histogram also shows a slight decrease in width with increasing $\lambda$.
Thus the degree of transport across the field is seen to have a very weak dependence on the value of the scattering mean free path.

SEPs are known to lose energy during propagation due to adiabatic deceleration \citep{Ruf1995, mas12}. Considering Figure~\ref{100MeV_3scat}, we find that the protons have final kinetic energies of $E=53, 64$ and $78$~MeV for the  increasing values of $\lambda$ presented, compared to their original injection energy $E_{0}=100$~MeV.

Figure~\ref{100MeV_3scat} demonstrates that protons originally injected at $E_{0}=100$~MeV are able to travel distances perpendicular to the field of the order 1~AU, on timescales typical of the duration of a gradual SEP event. 

\subsection{Drifts as the cause of displacement}
The perpendicular transport that is observed in Figure~\ref{100MeV_3scat} can be explained as due to magnetic field gradient (grad-$B$) and curvature drifts associated with the Parker spiral magnetic field.
Drift velocities in this large scale field can be calculated by means of standard single particle first-order adiabatic theory. \cite{bur68} first derived analytical expressions for Parker spiral drifts and, after making some assumptions and approximations, concluded that they are negligible for SEPs. 
In a concomitant paper \citep[][hereafter referred to as DMKL13]{anal_pap} we reconsidered drift velocities in the Parker spiral, calculating them in the $(\mathbf{e}_{l},\mathbf{e}_{\phi'},\mathbf{e}_{\theta'})$ coordinate system. This demonstrated analytically that drift velocities can be significant for SEPs.

Indicating the sum of  grad-$B$ and curvature drift velocities as $v_d$, its components for the magnetic field of Equation~(\ref{eqn_1}) is given by (DMKL13):
\begin{eqnarray} 
v_{d \phi'} & = &  \frac{m_0 \gamma \,  c}{q} \left( \frac{1}{2} \,v_{\perp}^2 - v_{\|}^2 \right) \, g(r,\theta) \label{eq.vdphipr}\\ 
v_{d \theta'} & = & - \frac{m_0 \gamma \,  c}{q} \left( \frac{1}{2} \,v_{\perp}^2 + v_{\|}^2 \right) \, f(r,\theta) \label{eq.vdthetapr}
\end{eqnarray}
where $v_{\|}$ and $v_{\perp}$ are the components of particle velocity parallel and perpendicular to the magnetic field and:
\begin{eqnarray}
g(r,\theta) &=&  \frac{a}{B_0 r_0^2} \, \frac{x^3 \cot{\theta}}{(x^2+1)^{3/2}} \label{eq.g_function} \\
f(r,\theta) &=& \frac{a}{B_0 r_0^2} \, \frac{x^2(x^2+2)}{(x^2+1)^2} \label{eq.f_function}
\end{eqnarray}
with $x$=$x(r,\theta)$=$r/a(\theta)$ and $a(\theta)$=$v_{sw}/(\Omega \sin{\theta})$. 
There is no drift along the magnetic field direction, i.e.~$v_{d l}$=0. 
The functions $g (r,\theta)$ and $f(r,\theta)$ describe the spatial variation of the $\phi'$ and $\theta'$ components of the grad-$B$ and curvature drifts respectively. 
While the function $f(r,\theta)$ is relatively constant in colatitude $\theta$,  $g(r,\theta)$ strongly depends on $\theta$ (see Figure 1 of DMKL13) and in particular it is zero at the heliographic equatorial plane.

In Equations~(\ref{eq.vdphipr}) and (\ref{eq.vdthetapr}), the term proportional to $v_{\perp}^2$ is due to grad-$B$ drift and the term proportional to $v_{\|}^2$ is due to curvature drift. When particles are injected at the Sun and propagate scatter-free, they are very quickly focussed to small pitch angles and only the second term, due to curvature drift, is nonzero. However, when scattering is present, particles will be characterised by a range of pitch angles and the first term, due to grad-$B$ drift, acquires a component of the velocity.

In the latitudinal $\theta'$ direction, the sign of the grad-$B$ and curvature drifts is the same (see Equation~\ref{eq.vdthetapr}).
Near the heliographic equatorial plane, the $\mathbf{e}_{\theta'}$ direction is approximately parallel to the solar rotation axis ($z$ axis);
for a positive ion at low latitude within the magnetic field of Equation~(\ref{eqn_1}) the drift motion is approximately anti-parallel to the $z$ axis.
The $\theta'$ drift is unidirectional regardless of scattering. The direction of this drift along the $\theta'$ axis is determined by the polarity of the magnetic field and particle charge alone. 

In the lateral $\phi'$ direction, the sign of the grad-$B$ and curvature drifts is opposite (see Equation~\ref{eq.vdphipr}); therefore, the history of scattering events (and resulting pitch-angle values) of a particle will affect the displacement magnitude and its direction relative to solar rotation.  
The $\phi'$ drift is bidirectional in the scattering case and essentially unidirectional without scattering. The drift direction and magnitude is dependent upon the magnetic field polarity, colatitude, difference between the particle perpendicular and parallel velocity $\frac{1}{2} v_{\perp}^2 - v_{\|}^2$, and particle charge.

Drifts explain all the observed features of the perpendicular transport in Figures~\ref{scat_free} and \ref{100MeV_3scat}. In the scatter-free case of Figure~\ref{scat_free}, the displacement in the negative $z$ direction is due to curvature drift. The reason why a drift in the $\phi'$ direction is not easily observed in Figure~\ref{scat_free} is that the function $g (r,\theta)$ appearing in Equation~\ref{eq.vdphipr} has a strong dependence on colatitude and is very small near the heliographic equatorial plane, and zero in the plane itself. In addition, for illustration purposes, the final simulation time for the scatter-free case (Figure~\ref{scat_free}) is shorter than in the scattering runs (Figure~\ref{100MeV_3scat}), therefore the latitudinal $\theta'$ drift has not forced the particles to drift far in colatitude from the equatorial plane.

The displacements in $\theta'$ and $\phi'$ seen in the scattering cases of Figure~\ref{100MeV_3scat} are due to a combination of grad-$B$ and curvature drifts. 
The drifts force particles away from the equatorial plane towards regions in colatitude where the function $g (r,\theta)$ is nonzero and the $\phi'$ drift can become significant, giving rise to the histograms of Figure 2.

\begin{figure}[!t]
\epsscale{1.0}
%\epsscale{0.9}
\centering
%\rotate
\plotone{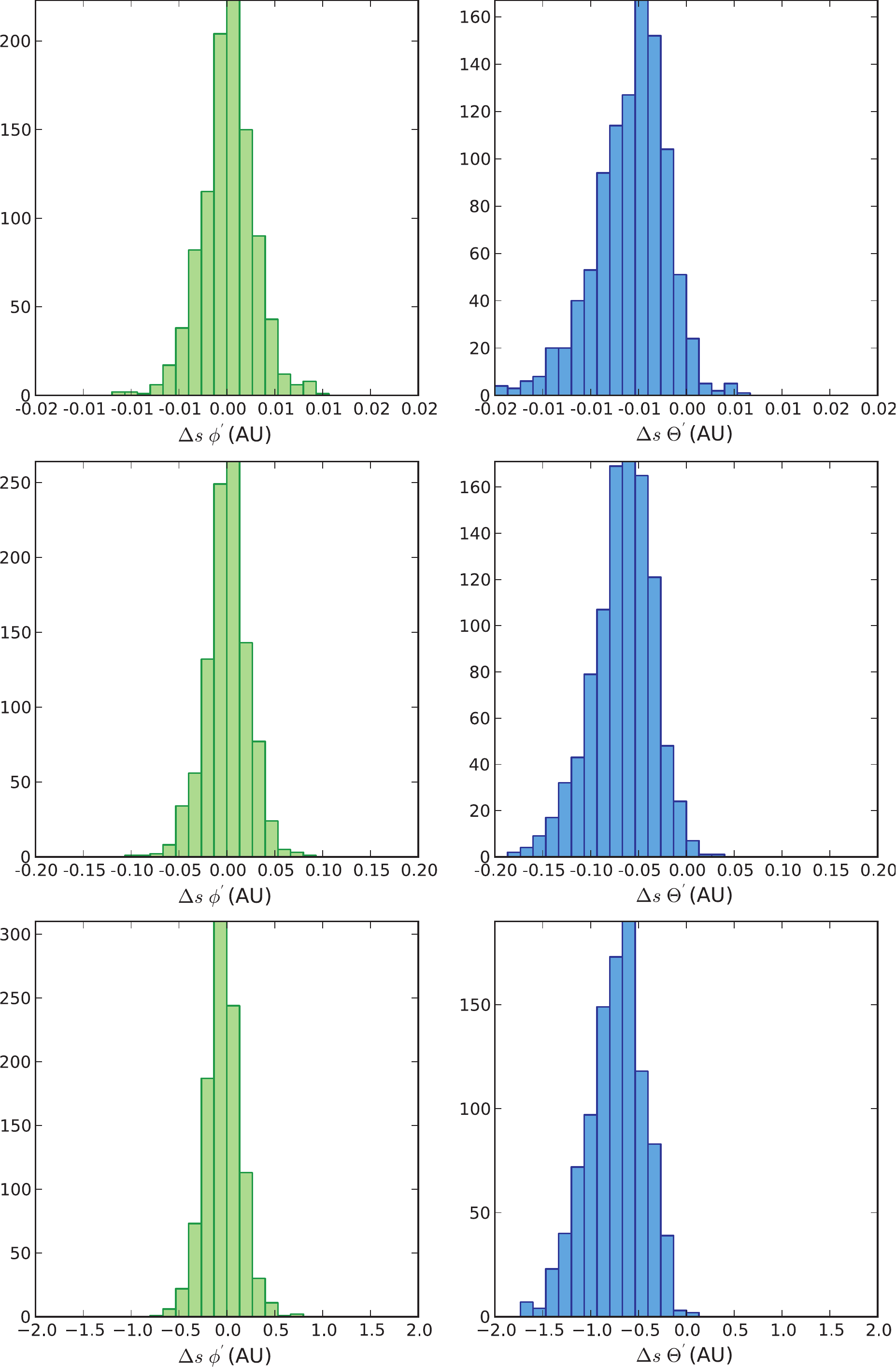}
\caption{Histograms of lateral displacement $\Delta s_{\phi'}$ (left) and latitudinal displacement $\Delta s_{\theta'}$ (right) for proton energy $E_0$=1~MeV (top row), 10~MeV (middle row), and 100~MeV (bottom row). The scattering mean free path is $\lambda=1$~AU, $\delta_{0}$=0, $N$=1000 and $t$=4~days. Note the change in drift displacement scale on the histograms for each energy.}
\label{E_dependence}
\end{figure}

There is little difference between the distributions of lateral displacement $\Delta s_{\phi'}$ shown in the histograms of Figure~\ref{100MeV_3scat}, considering that the mean free path varies by around two orders of magnitude. This is contrary to what one might expect, since the displacement in the $\phi'$ direction is dependent upon the difference between $v_{\perp}^2$ and $v_{\|}^2$ (Equation~\ref{eq.vdphipr}); considering the balance between scattering and focussing of the pitch angle as the particles propagate, we may expect $\Delta s_{\phi'}$ to be strongly dependent on mean free path.  
The reason why the dependence is weak is because a particle that scatters to e.g. pitch angle $\alpha$=90$^{\circ}$ at large distances from the Sun will take much longer to focus than a particle that is injected close to the Sun with the same pitch angle (see Figure 4 of DMKL13). This means that, once established, the population with large perpendicular velocities will persist over time. Hence even a low level of scattering will produce considerable grad-B drift.

It is important to note that the $\phi'$ drift seen in Figure~\ref{100MeV_3scat}, where the injection region is centered at the heliographic equatorial plane, is essentially a lower limit due to the dependence of $g(r,\theta)$ on colatitude. The dependence of drift on colatitude will be further discussed in Section \ref{sec.latitude}.

\subsection{Energy dependence of drift}\label{sec.energy}
% Mean final energy corresponding to plot E_0=1 MeV E_f=0.4 MeV, E_0=10 MeV E_f=5 MeV, E_0=100 MeV E_f=64 MeV

To study how the amount of drift depends on particle energy, simulations for monoenergetic populations of 1000 protons were carried out for $E_0$=1, 10 and 100 MeV, with a mean free path $\lambda$=1 AU and injection at $\delta_{0}$=0.
Figure~\ref{E_dependence} shows the dependence of drift on proton energy at $t$=4~days (note that the abscissa scale is different at each energy).
Displacements increase by approximately a factor of 10 as the proton injection energy increases from 1 to 10~MeV and from 10 to 100~MeV. This is expected due to the dependence of the magnitude of drift velocities on 1/2$\gamma m_0 \, v_{\perp}^2$ and $\gamma m_0 \, v_{\|}^2$ 
(see Equations~\ref{eq.vdphipr} and \ref{eq.vdthetapr}). 

\subsection{Latitude dependence of drift}\label{sec.latitude}
% Mean final energy of plots in Fig. Lat=20 E=64 MeV, Lat=40 E=66, Lat=60 E=70.

\begin{figure*}[!ht]
%\epsscale{1.0}
%\epsscale{0.9}
%\centering
%\rotate
\plotone{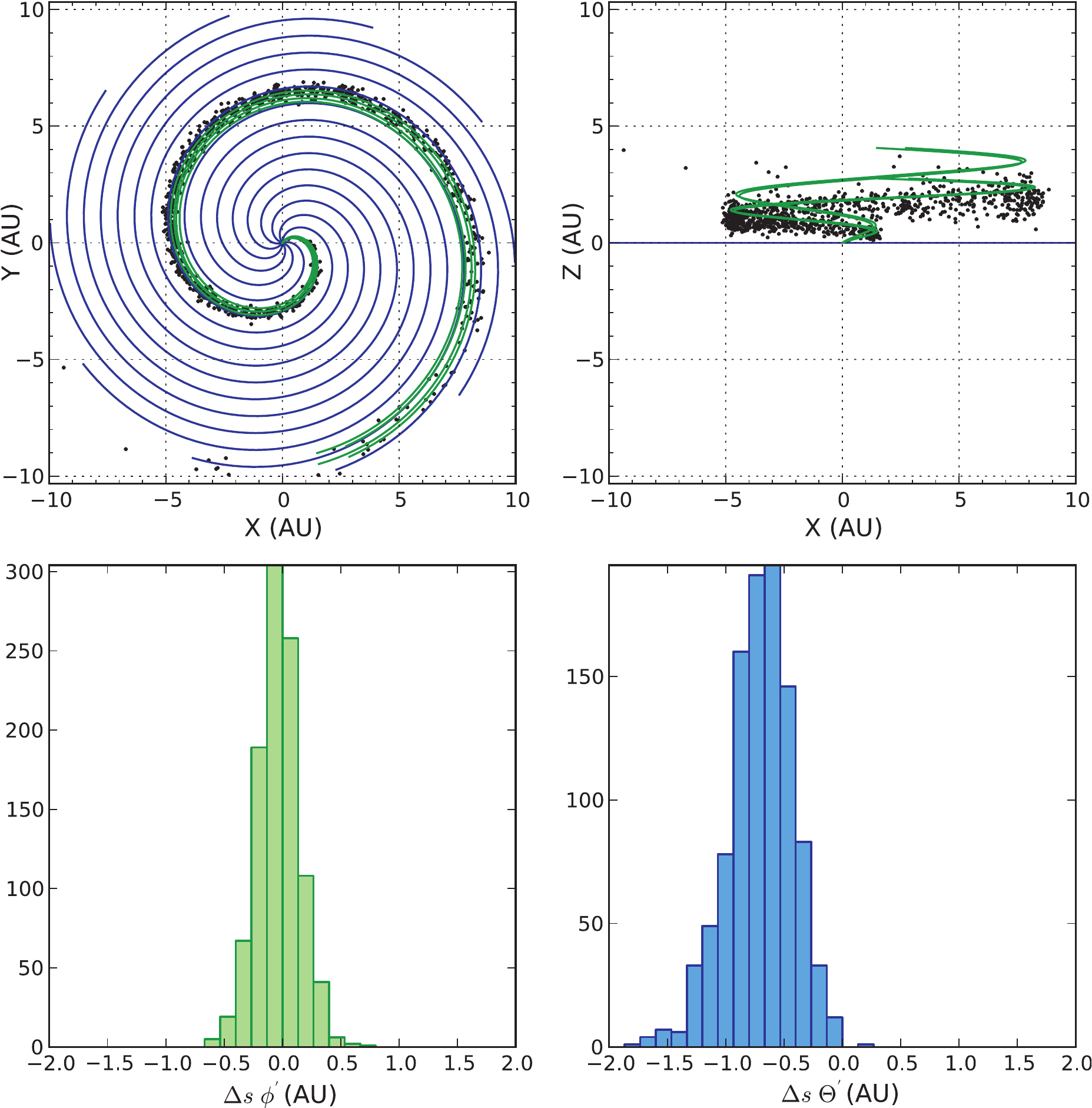}
\plotone{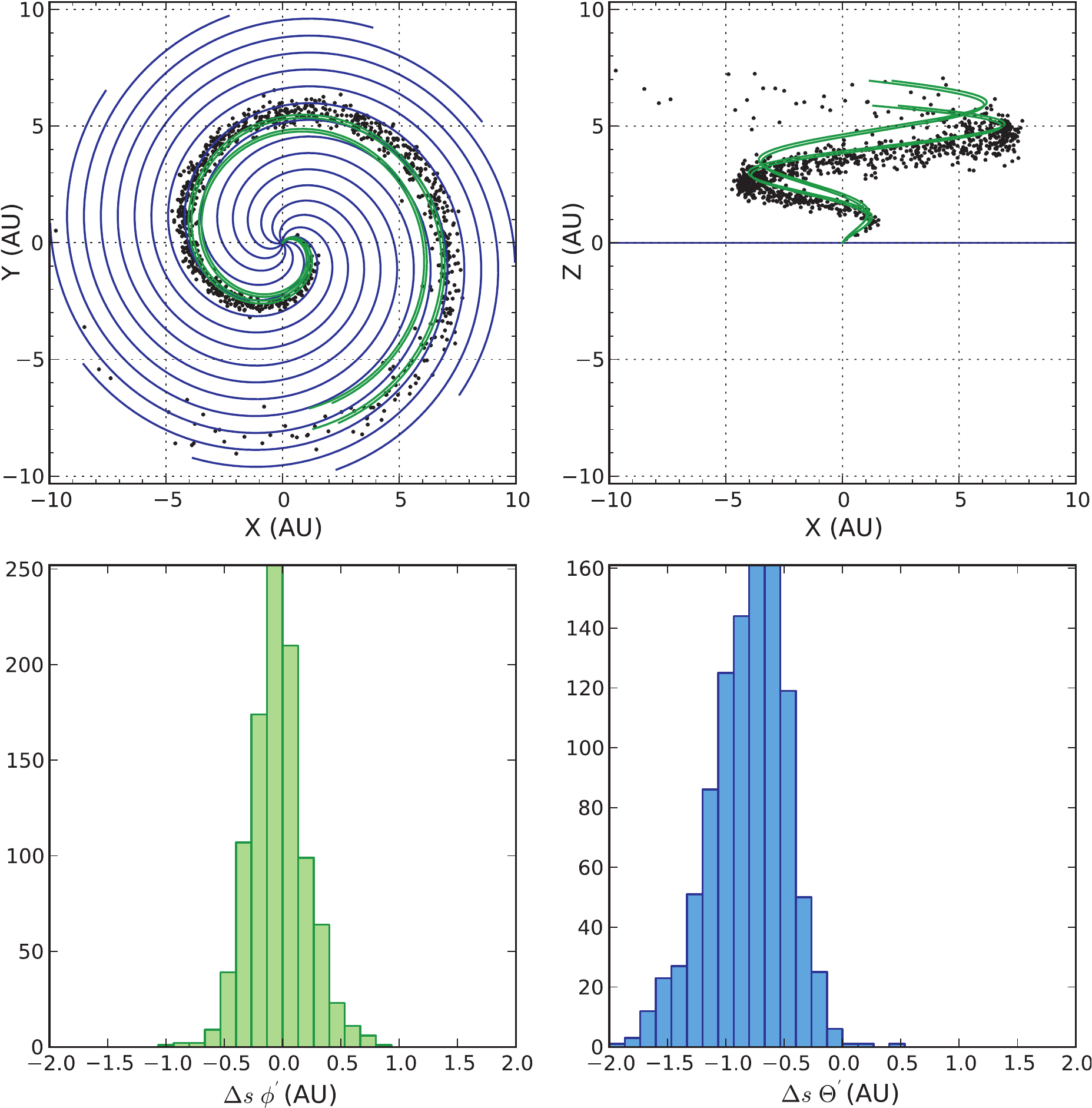}
\plotone{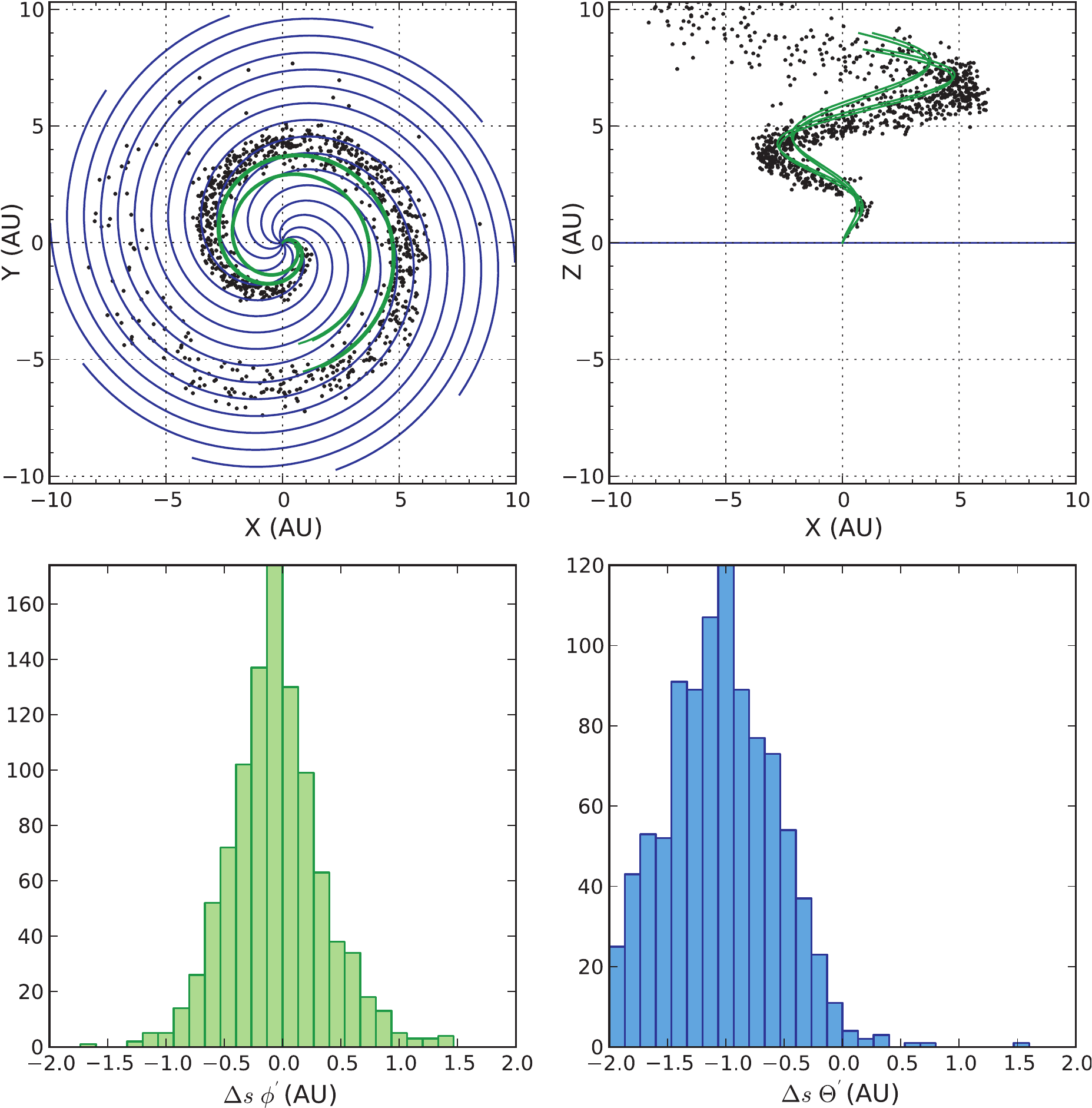}
\caption{Particle locations and displacement histograms for injection at heliographic latitudes $\delta_{0}$=20 (top), 40 (middle) and 60 (bottom), for protons with $E_0$=100 MeV, $\lambda=1$~AU and $N$=1000. In the location plots, the blue curves show the equatorial IMF with the field lines bounding the particle injection region in green, as in Figure~\ref{100MeV_3scat}.  The online edition of the paper includes a movie showing the $\delta_{0}$=40 case.}
\label{lat_dependence}
\end{figure*}

\begin{figure}[!t]
%\epsscale{1.0}
%\epsscale{0.9}
\centering
%\rotate
\plotone{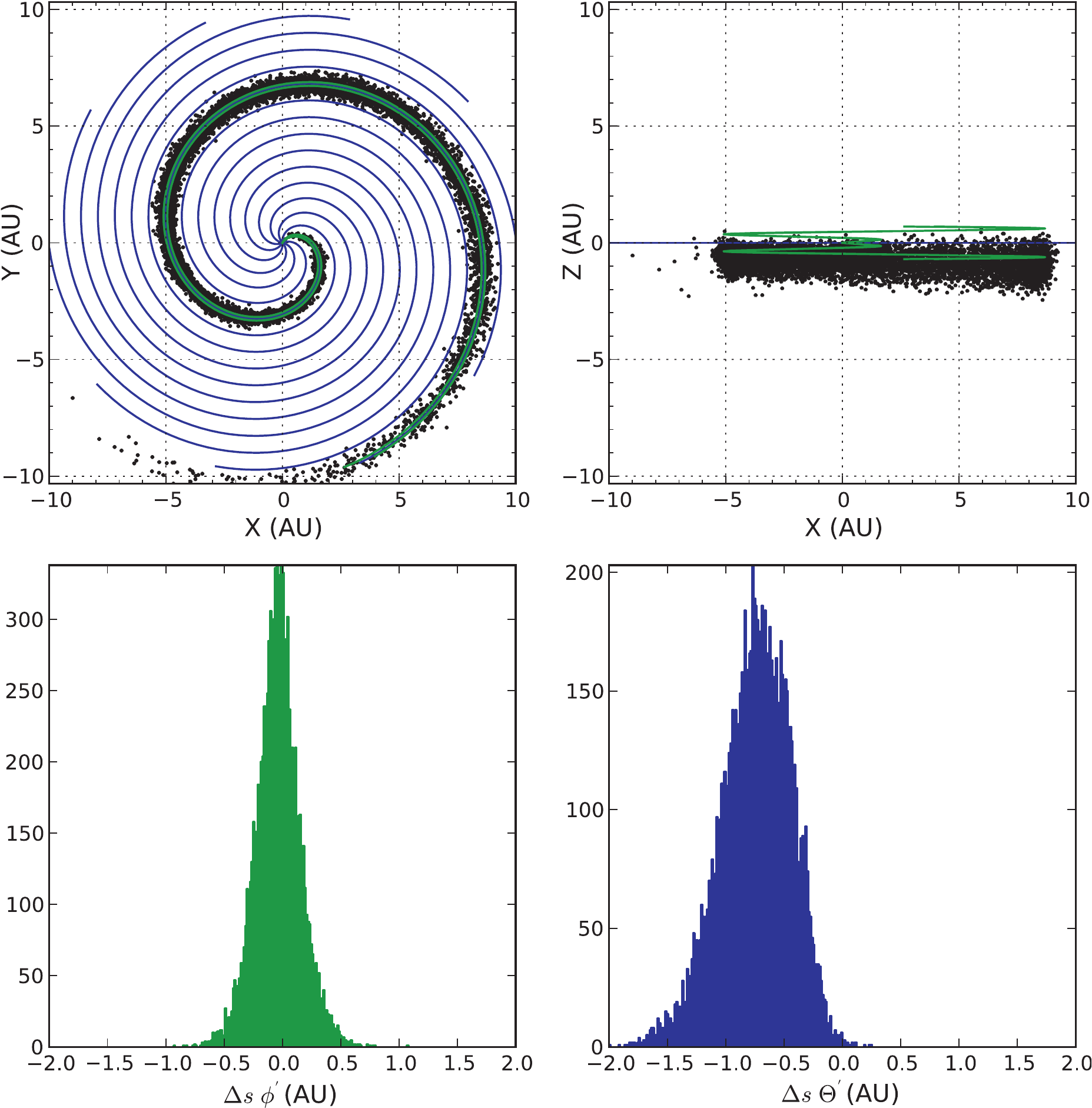}
\caption{Particle locations with lateral and latitudinal displacement histograms for 10,000 protons at $E_0$=100~MeV, $t$=4 days, $\lambda$=1~AU, and $\delta_{0}$=0 (c.f. center panels of Figure~\ref{100MeV_3scat}).  The online edition of the paper includes a movie to accompany the figure.}
\label{large_n}
\end{figure}

Figure~\ref{lat_dependence} shows the final locations and displacement histograms for $E_0$=100~MeV protons injected at three heliospheric latitudes: $\delta_{0}$=20, 40 and 60. The scattering mean free path is $\lambda$=1 AU, so that these results compare with the $\delta_{0}$=0 run shown in the middle column of Figure~\ref{100MeV_3scat}.  

It should be noted that the drift velocity vectors are perpendicular to the local magnetic field at the location of a particle.
In the case of Figure~\ref{100MeV_3scat}, where the particles are located around the equatorial plane, the lateral $\phi'$ and latitudinal $\theta'$ drift velocity vectors lie approximately in the ($x$-$y$) and ($x$-$z$) planes respectively. This allows a visualisation of the particle drift displacements, along $\phi'$ and $\theta'$, from the particle locations projected in ($x$-$y$) and ($x$-$z$).
In contrast, considering Figure~\ref{lat_dependence}, when particles are located at non-zero latitudes, the displacements are not parallel to the ($x$-$y$) and ($x$-$z$) planes, therefore the contribution of the $\phi'$ and $\theta'$ components of drift are confounded in the ($x$-$y$) and ($x$-$z$) projections. 

For example, in Figure~\ref{lat_dependence}, the particles appear to show an
asymmetric sideways drift from the injection region field lines in the ($x$-$y$) projection, but the histogram for $\phi'$ shows that the displacement is approximately symmetric about zero. This is
because the $\theta'$ drift is perpendicular to the surface of the cone described by the Parker spiral magnetic
field, with a negative $z$ component for positive charge and magnetic field polarity, and not parallel to the $z$ axis; therefore, when viewed in the ($x$-$y$) projection there appears to be a shift of the particle locations
outwards of the spiral for positive latitudes and inward of the spiral for negative latitudes, due to the $\theta'$ drift having a component in the ($x$-$y$) plane. 
The $\phi'$ drift also has an
equivalent projection effect when viewed in projection. This demonstrates that the histograms are important to quantify 
the contribution of drift in the $\theta'$ and $\phi'$ directions.

The histograms in Figure~\ref{lat_dependence} show that the width of the displacement distributions increase with injection latitude. The peak of the $\Delta s_{\theta'}$ histogram also becomes increasingly negative with latitude. The trend of increasing drift with latitude is consistent with the analytical expressions for drift velocities (DMKL13).

\subsection{Varying the number of simulated particles}\label{sec.nvalue}

Figure~\ref{large_n} shows the same simulation as displayed in the middle panel of Figure~\ref{100MeV_3scat}, but with an increased number of test particles, $N$=10000.
It can be seen that the displacement histograms correspond well. This suggests that a simulation of $N$=1000 particles is sufficient to resolve the distribution of displacements due to drifts. 

\subsection{Heavy ions}\label{sec.heavyions}
% mean final energy q=20 E_f=64 Mev/nuc, q=15 E_f=64 MeV/nuc

The magnitude of the drift velocities is dependent upon the particle mass to charge ratio $m_{0} \gamma/q$ (see Equations~\ref{eq.vdphipr} and \ref{eq.vdthetapr}). Since SEP heavy ions are typically not fully ionized, i.e. have low q-values \citep{kle07}, they undergo correspondingly larger drifts. 

\begin{figure}[!t]
\plotone{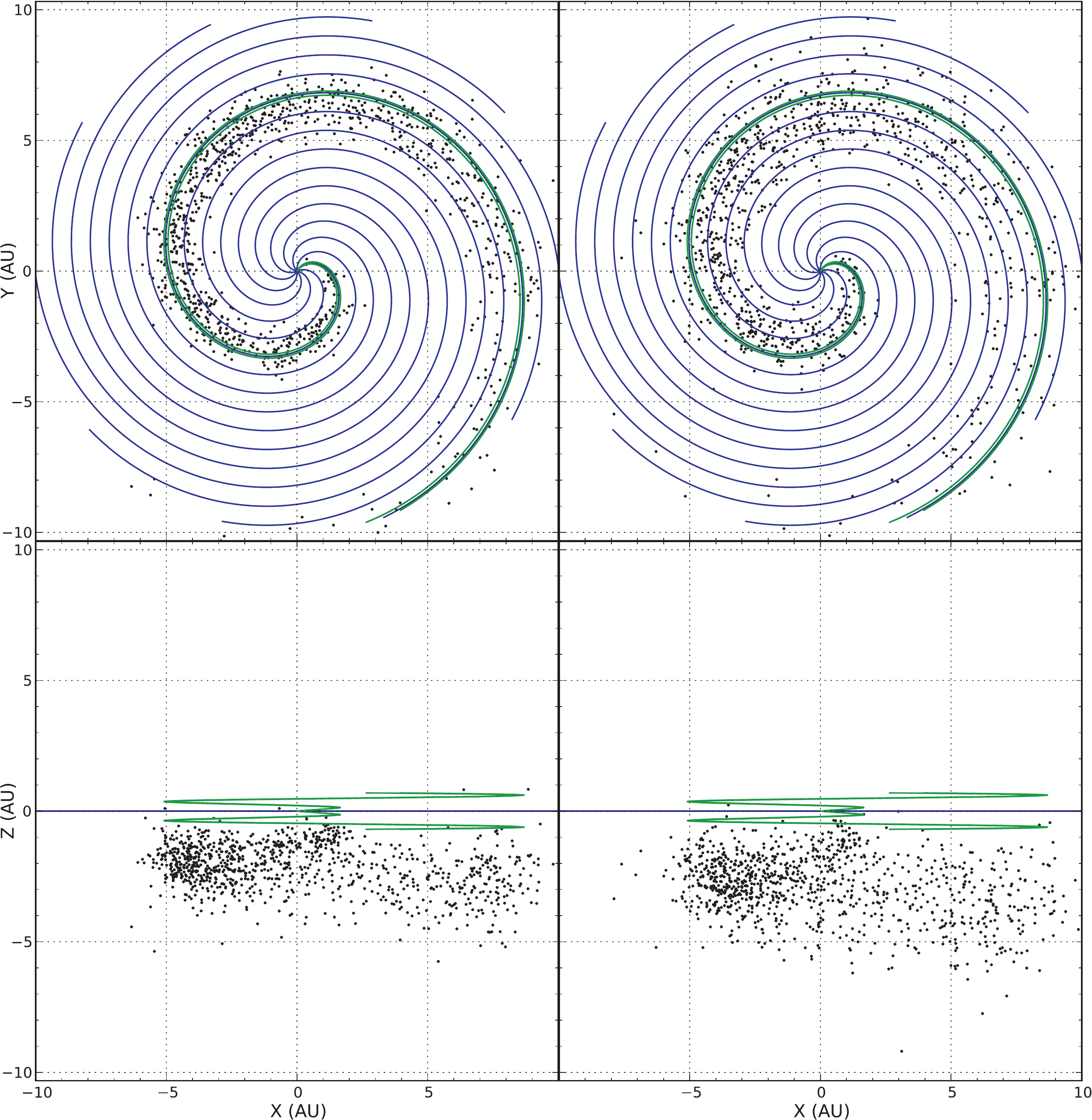}
\caption{Locations of 1000 Fe ions with $E_0$=100~MeV/nuc for injection at $\delta_{0}$=0, $\lambda$=1~AU, and $t$=4 days.  The results are shown for ionization state $^{56}$Fe$^{20+}$ (left) and $^{56}$Fe$^{15+}$ (right). The online edition of the paper includes movies of both ionization states.}
\label{fe_runs}
\end{figure}

Figure~\ref{fe_runs} shows the location of Fe ions, with injection at $\delta_{0}=0$, $\lambda=1$~AU and $t$=4 days, for ionization states of $^{56}$Fe$^{20+}$ (left) and $^{56}$Fe$^{15+}$ (right), and energy $E_0$=100 MeV/nuc, i.e. ions of the same speed as 100 MeV protons, allowing comparison to the centre panel of Figure~\ref{100MeV_3scat}. An ionization state of 15 represents a typical charge state for SEP ions in gradual events, while 20 is typical of impulsive events \citep{rea99}, although recent observations indicate the charge state separation of the two event types may not be so clear \citep{kle07}.

The amount of drift experienced by the iron ions is significantly larger than for protons, with the $^{56}$Fe$^{15+}$ experiencing the largest drift. It is not possible to calculate the displacement histograms for the large iron drifts presented in Figure~\ref{fe_runs}, since due to the large displacements the numerical solution of the equation for the target point is problematic. However, we may estimate that the mean drift is on the order of 2 and 3~AU for the $^{56}$Fe$^{20+}$ and $^{56}$Fe$^{15+}$ ions respectively. 

Note that the ($x$-$y$) projection gives the impression that the $\phi'$ drift is enhanced particularly comparing between different charge states, but it is not possible to distinguish the contribution of the drift components in this projection as described in Section~\ref{sec.latitude}. The increased $\theta'$ drift in the $^{56}$Fe$^{15+}$ case gives rise to an increased asymmetry inward of the spiral in the ($x$-$y$) projection in comparison to $^{56}$Fe$^{20+}$. 

\section{Discussion and Conclusions} \label{sec.conclusions}
Simulations of the propagation of SEPs injected from a localised region at the Sun were carried out by means of a full-orbit test particle code, for a Parker spiral IMF.
The main results obtained are as follows:
\begin{enumerate}
\item Significant drift across the magnetic field is seen for SEP ions, due to the curvature and gradient of the Parker spiral IMF. 
Both the lateral $\phi'$ and latitudinal $\theta'$ components of the drift are nonzero with or without the presence of scattering. 
Protons injected at 100~MeV have an average displacement from the original field line of $\sim$1~AU, over a simulation time of 4~days.

\item The amount of drift strongly increases with increasing absolute value of heliographic latitude. 

\item Partially ionized SEP heavy ions experience larger drifts than protons of the same speed, as drift velocities depend on the $m_{0} \gamma/q$ ratio. $^{56}$Fe$^{15+}$ ions injected at 100~MeV/nuc show an average displacement of $\sim$3~AU after a simulation time of 4~days. 

\end{enumerate}

Our results show that drifts are important in the propagation of high energy SEPs and contribute to their transport across the magnetic field. Drifts need to be included in transport models of energetic SEP ions especially for large $m_{0} \gamma/q$.  
The amount and characteristics of drift seen in the simulations is consistent with analytical calculations of single particle first order drifts from adiabatic theory \citep{anal_pap}.
The direction of the drifts for SEPs will follow the well known GCR pattern, for a standard heliospheric A$^{+}$ or A$^{-}$ IMF configuration \citep{Jok1977}.
The drifts associated with the large-scale Parker spiral fields that are studied in this paper are likely to be a lower limit to the amount of drift-induced cross-field transport for SEPs. It is expected that large magnetic field gradients and structures with small radius of curvature, e.g. associated with turbulence in the IMF, will produce additional drift.

Scattering modifies the distributions of drift displacement when compared to the scatter-free case, however in the scattering case the magnitude of drift is very weakly dependent on the level of scattering.
In the scatter free case, drift is due solely to the curvature drift. Drifts are significant even at low levels of scattering. The scatter free case gives the maximum limit of drift displacement magnitude. 

The magnitude of drift increases with particle energy. While the initial energy of the particles injected in our simulations is relatively high, deceleration during propagation reduces their energy \citep{mas12} so that they contribute to fluxes at energies lower than their injection energy $E_{0}$. This allows the possibility that large drifts experienced by particles at high energies can cascade down to particles at lower energies.
Determination of the drift magnitude is a multivariate problem, due to the complex dependence on energy, latitude, mass to charge ratio, distance from the Sun, and time since the start of the event.  It is therefore not possible to define a simple threshold in terms of particle energy and species above which significant drift occurs. The results presented here serve to illustrate the effect of drift for SEPs, and that in general it is non-negligible.

Simulations carried out for protons at energies in the GeV range, which can cause terrestrial Ground Level Enhancement (GLE) events \citep{gop12}, show an even stronger effect of drifts and will be discussed in a future publication.  

Concerning heavy ions, it is known that the mass over charge ratio orders a variety of heavy ion characteristics in SEP events \citep[e.g.][]{rea99, mas12}. The dependence of drift velocities on $m_{0} \gamma/q$ is another possible mechanism that may cause species dependent behaviour, which has never been consider up to the present time, such as fractionation of the abundances, and is likely to play an important role.

The strong dependence of drifts on heliographic latitude may help to explain the SEP observations made by Ulysses at high heliolatitudes \citep{mck03}. In particular the strong drifts can explain the fact that particles were able to reach the spacecraft independently of the heliolongitude of the source and that the separation in heliolatitude between spacecraft and source region was the parameter that ordered the characteristics of the observations best \citep{dal03a, dal03b}. Our simulations show that drifts are effective in generating a widely spread population of energetic particles in the heliosphere \citep[c.f.][]{mck72}, which could potentially be reaccelerated. 

Inclusion of drift effects into transport models will be possible by means of a drift term which, unlike for the case for GCRs, needs to include a dependence on the particle pitch-angle. The analytical expressions for drift velocities derived in \cite{anal_pap} may form the basis of such a description assuming a Parker spiral field. The cross field transport associated with drift is not symmetric in the latitudinal direction and cannot be incorporated into models by means of a symmetric perpendicular diffusion coefficient. 

Our results show for the first time that drift-induced transport across the field cannot be neglected in the analysis and modelling of the fluxes of high energy SEP protons and heavy ions.

\acknowledgments
This work has received funding from the European Union Seventh
Framework Programme (FP7/2007-2013) under grant agreement n.~263252
[COMESEP]. TL acknowledges support from the UK Science and Technology Facilities Council (STFC) (grant ST/J001341/1).

\bibliographystyle{apj}
\bibliography{ms}

\newcommand{\cds}{{\itshape CDS Software Note}}\newcommand{\pc}{{\itshape
  Personal communication}}\newcommand{\sub}{{\itshape
  Submitted}}\newcommand{\inprep}{{\itshape In
  preparation}}\newcommand{\RSPSA}{Proc. R. Soc. Lond.
  A}\newcommand{\RvMP}{Reviews of Modern Physics}
\begin{thebibliography}{31}
\expandafter\ifx\csname natexlab\endcsname\relax\def\natexlab#1{#1}\fi

\bibitem[{{Aran} {et~al.}(2006){Aran}, {Sanahuja}, \& {Lario}}]{Ara2006}
{Aran}, A., {Sanahuja}, B., \& {Lario}, D. 2006, Advances in Space Research,
  37, 1240

\bibitem[{{Burns} \& {Halpern}(1968)}]{bur68}
{Burns}, J.~A., \& {Halpern}, G. 1968, \jgr, 73, 7377

\bibitem[{{Dalla} {et~al.}(2003{\natexlab{a}}){Dalla}, {Balogh}, {Krucker},
  {Posner}, {M{\"u}ller-Mellin}, {Anglin}, {Hofer}, {Marsden}, {Sanderson},
  {Heber}, {Zhang}, \& {McKibben}}]{dal03a}
{Dalla}, S., {Balogh}, A., {Krucker}, S., {Posner}, A., {M{\"u}ller-Mellin},
  R., {Anglin}, J.~D., {Hofer}, M.~Y., {Marsden}, R.~G., {Sanderson}, T.~R.,
  {Heber}, B., {Zhang}, M., \& {McKibben}, R.~B. 2003{\natexlab{a}}, Annales
  Geophysicae, 21, 1367

\bibitem[{{Dalla} {et~al.}(2003{\natexlab{b}}){Dalla}, {Balogh}, {Krucker},
  {Posner}, {M{\"u}ller-Mellin}, {Anglin}, {Hofer}, {Marsden}, {Sanderson},
  {Tranquille}, {Heber}, {Zhang}, \& {McKibben}}]{dal03b}
{Dalla}, S., {Balogh}, A., {Krucker}, S., {Posner}, A., {M{\"u}ller-Mellin},
  R., {Anglin}, J.~D., {Hofer}, M.~Y., {Marsden}, R.~G., {Sanderson}, T.~R.,
  {Tranquille}, C., {Heber}, B., {Zhang}, M., \& {McKibben}, R.~B.
  2003{\natexlab{b}}, \grl, 30, 8035

\bibitem[{{Dalla} \& {Browning}(2005)}]{dall05}
{Dalla}, S., \& {Browning}, P.~K. 2005, \aap, 436, 1103

\bibitem[{{Dalla} {et~al.}(2013){Dalla}, {Marsh}, {Kelly}, \&
  {Laitinen}}]{anal_pap}
{Dalla}, S., {Marsh}, M.~S., {Kelly}, J., \& {Laitinen}, T. 2013, \jgr, \sub, arXiv:1307.2165

\bibitem[{{Dresing} {et~al.}(2012){Dresing}, {G{\'o}mez-Herrero}, {Klassen},
  {Heber}, {Kartavykh}, \& {Dr{\"o}ge}}]{Dre2012}
{Dresing}, N., {G{\'o}mez-Herrero}, R., {Klassen}, A., {Heber}, B.,
  {Kartavykh}, Y., \& {Dr{\"o}ge}, W. 2012, \solphys, 281, 281

\bibitem[{{Dr{\"o}ge} {et~al.}(2010){Dr{\"o}ge}, {Kartavykh}, {Klecker}, \&
  {Kovaltsov}}]{Dro2010}
{Dr{\"o}ge}, W., {Kartavykh}, Y.~Y., {Klecker}, B., \& {Kovaltsov}, G.~A. 2010,
  \apj, 709, 912

\bibitem[{{Ferreira} \& {Potgieter}(2004)}]{Fer2004}
{Ferreira}, S.~E.~S., \& {Potgieter}, M.~S. 2004, \apj, 603, 744

\bibitem[{{Gopalswamy} {et~al.}(2012){Gopalswamy}, {Xie}, {Yashiro}, {Akiyama},
  {M{\"a}kel{\"a}}, \& {Usoskin}}]{gop12}
{Gopalswamy}, N., {Xie}, H., {Yashiro}, S., {Akiyama}, S., {M{\"a}kel{\"a}},
  P., \& {Usoskin}, I.~G. 2012, \ssr, 171, 23

\bibitem[{{Jokipii} {et~al.}(1977){Jokipii}, {Levy}, \& {Hubbard}}]{Jok1977}
{Jokipii}, J.~R., {Levy}, E.~H., \& {Hubbard}, W.~B. 1977, \apj, 213, 861

\bibitem[{{Kallenrode}(1997)}]{kal97}
{Kallenrode}, M.-B. 1997, \jgr, 102, 22335

\bibitem[{{Kelly} {et~al.}(2012){Kelly}, {Dalla}, \& {Laitinen}}]{kel12}
{Kelly}, J., {Dalla}, S., \& {Laitinen}, T. 2012, \apj, 750, 47

\bibitem[{{Klecker} {et~al.}(2007){Klecker}, {M{\"o}bius}, \&
  {Popecki}}]{kle07}
{Klecker}, B., {M{\"o}bius}, E., \& {Popecki}, M.~A. 2007, \ssr, 130, 273

\bibitem[{{Laitinen} {et~al.}(2012){Laitinen}, {Dalla}, \& {Kelly}}]{lai12}
{Laitinen}, T., {Dalla}, S., \& {Kelly}, J. 2012, \apj, 749, 103

\bibitem[{{Lario} {et~al.}(1998){Lario}, {Sanahuja}, \& {Heras}}]{Lar1998}
{Lario}, D., {Sanahuja}, B., \& {Heras}, A.~M. 1998, \apj, 509, 415

\bibitem[{{Luhmann} {et~al.}(2010){Luhmann}, {Ledvina}, {Odstrcil}, {Owens},
  {Zhao}, {Liu}, \& {Riley}}]{Luh2010}
{Luhmann}, J.~G., {Ledvina}, S.~A., {Odstrcil}, D., {Owens}, M.~J., {Zhao},
  X.-P., {Liu}, Y., \& {Riley}, P. 2010, Advances in Space Research, 46, 1

\bibitem[{{Mason} {et~al.}(2012){Mason}, {Li}, {Cohen}, {Desai}, {Haggerty},
  {Leske}, {Mewaldt}, \& {Zank}}]{mas12}
{Mason}, G.~M., {Li}, G., {Cohen}, C.~M.~S., {Desai}, M.~I., {Haggerty}, D.~K.,
  {Leske}, R.~A., {Mewaldt}, R.~A., \& {Zank}, G.~P. 2012, \apj, 761, 104

\bibitem[{{McKibben}(1972)}]{mck72}
{McKibben}, R.~B. 1972, \jgr, 77, 3957

\bibitem[{{McKibben} {et~al.}(2003){McKibben}, {Connell}, {Lopate}, {Zhang},
  {Anglin}, {Balogh}, {Dalla}, {Sanderson}, {Marsden}, {Hofer}, {Kunow},
  {Posner}, \& {Heber}}]{mck03}
{McKibben}, R.~B., {Connell}, J.~J., {Lopate}, C., {Zhang}, M., {Anglin},
  J.~D., {Balogh}, A., {Dalla}, S., {Sanderson}, T.~R., {Marsden}, R.~G.,
  {Hofer}, M.~Y., {Kunow}, H., {Posner}, A., \& {Heber}, B. 2003, Annales
  Geophysicae, 21, 1217

\bibitem[{{Pei} {et~al.}(2006){Pei}, {Jokipii}, \& {Giacalone}}]{Pei2006}
{Pei}, C., {Jokipii}, J.~R., \& {Giacalone}, J. 2006, Astrophysical Journal,
  641, 1222

\bibitem[{Press {et~al.}(1996)Press, Teukolsky, Vetterling, \&
  Flannery}]{pre96}
Press, W.~H., Teukolsky, S.~A., Vetterling, W.~T., \& Flannery, B.~P. 1996,
  Numerical recipes in Fortran 90 (2nd ed.): the art of parallel scientific
  computing (New York, NY, USA: Cambridge University Press)

\bibitem[{{Reames}(1999)}]{rea99}
{Reames}, D.~V. 1999, \ssr, 90, 413

\bibitem[{{Roelof}(1969)}]{Roe1969}
{Roelof}, E.~C. 1969, in Lectures in High-Energy Astrophysics, ed.
  H.~{{\"O}gelman} \& J.~R. {Wayland}, 111

\bibitem[{{Ruffolo}(1995)}]{Ruf1995}
{Ruffolo}, D. 1995, \apj, 442, 861

\bibitem[{{Schwadron} {et~al.}(2010){Schwadron}, {Townsend}, {Kozarev},
  {Dayeh}, {Cucinotta}, {Desai}, {Golightly}, {Hassler}, {Hatcher}, {Kim},
  {Posner}, {PourArsalan}, {Spence}, \& {Squier}}]{Sch2010}
{Schwadron}, N.~A., {Townsend}, L., {Kozarev}, K., {Dayeh}, M.~A., {Cucinotta},
  F., {Desai}, M., {Golightly}, M., {Hassler}, D., {Hatcher}, R., {Kim}, M.-Y.,
  {Posner}, A., {PourArsalan}, M., {Spence}, H.~E., \& {Squier}, R.~K. 2010,
  Space Weather, 8, 0

\bibitem[{{Tautz} {et~al.}(2011){Tautz}, {Shalchi}, \& {Dosch}}]{tau11}
{Tautz}, R.~C., {Shalchi}, A., \& {Dosch}, A. 2011, Journal of Geophysical
  Research (Space Physics), 116, 2102

\bibitem[{{Wang} {et~al.}(2012){Wang}, {Qin}, \& {Zhang}}]{Wan2012}
{Wang}, Y., {Qin}, G., \& {Zhang}, M. 2012, \apj, 752, 37

\bibitem[{{Wiedenbeck} {et~al.}(2013){Wiedenbeck}, {Mason}, {Cohen}, {Nitta},
  {G{\'o}mez-Herrero}, \& {Haggerty}}]{wie13}
{Wiedenbeck}, M.~E., {Mason}, G.~M., {Cohen}, C.~M.~S., {Nitta}, N.~V.,
  {G{\'o}mez-Herrero}, R., \& {Haggerty}, D.~K. 2013, \apj, 762, 54

\bibitem[{{Winge} \& {Coleman}(1968)}]{Win1968}
{Winge}, C.~R.~J., \& {Coleman}, P.~J.~J. 1968, \jgr, 73, 165

\bibitem[{{Zhang} {et~al.}(2009){Zhang}, {Qin}, \& {Rassoul}}]{Zha2009}
{Zhang}, M., {Qin}, G., \& {Rassoul}, H. 2009, \apj, 692, 109

\end{thebibliography}

\end{document}